\documentclass[times,twocolumn,final]{elsarticle}

\usepackage{medima}
\usepackage{framed,multirow}
\usepackage{graphicx}
\usepackage{subfigure}
\usepackage{amsmath}
\usepackage{multirow}
\usepackage{comment}
\usepackage{color}
\usepackage{ulem}
\usepackage{booktabs}
\usepackage{dsfont}
\usepackage{amssymb}
\usepackage{latexsym}

\usepackage{url}
\usepackage{xcolor}

\usepackage{hyperref}

\newcommand{\etc}{\mbox{etc.\ }}
\newcommand{\eg}{\mbox{e.g.,\ }}

\newcommand{\ie}{\mbox{i.e.,\ }}
\newcommand{\aka}{\mbox{a.k.a.\ }}

\definecolor{newcolor}{rgb}{.8,.349,.1}

\journal{Medical Image Analysis}

\begin{document}

\verso{Jiayu Huo \textit{et~al.}}

\begin{frontmatter}

\title{Self-supervised Brain Lesion Generation for Effective Data Augmentation of Medical Images}%

\author[1]{Jiayu \snm{Huo}\corref{cor1}}
\author[1]{S\'{e}bastien \snm{Ourselin}}
\author[1]{Rachel \snm{Sparks}}

\cortext[cor1]{Corresponding author: Jiayu Huo (\href{mailto:jiayu.huo@kcl.ac.uk}{\textsc{jiayu.huo@kcl.ac.uk}})}

\address[1]{School of Biomedical Engineering and Imaging Sciences (BMEIS), King's College London, London, UK.}

\received{1 May 2013}
\finalform{10 May 2013}
\accepted{13 May 2013}
\availableonline{15 May 2013}
\communicated{S. Sarkar}

\begin{abstract}
Accurate brain lesion delineation is important for planning neurosurgical treatment. Automatic brain lesion segmentation methods based on convolutional neural networks have demonstrated remarkable performance. However, neural network performance is constrained by the lack of large-scale well-annotated training datasets. In this manuscript, we propose a comprehensive framework to efficiently generate new samples for training a brain lesion segmentation model. We first train a lesion generator, based on an adversarial autoencoder, in a self-supervised manner. Next, we utilize a novel image composition algorithm, Soft Poisson Blending, to seamlessly combine synthetic lesions and brain images to obtain training samples. Finally, to effectively train the brain lesion segmentation model with augmented images we introduce a new prototype consistence regularization to align real and synthetic features. Our framework is validated by extensive experiments on two public brain lesion segmentation datasets: ATLAS v2.0 and Shift MS. Our method outperforms existing brain image data augmentation schemes. For instance, our method improves the Dice from 50.36\% to 60.23\% compared to the U-Net with conventional data augmentation techniques for the ATLAS v2.0 dataset.
\end{abstract}


\end{frontmatter}


\section{Introduction}\label{sec:introduction}
{B}{rain} lesions are often indicative of serious neurological conditions, from cancer to stroke. Magnetic Resonance (MR) Imaging is widely used to detect brain lesions as MR provides excellent soft-tissue contrast, allowing for a clear distinction between healthy and abnormal brain tissue. Accurate segmentation of brain lesions is crucial for quantitative analysis of lesion progression and planning surgical treatments. The current clinical standard is human delineation of the brain lesion boundary by an expert which is tedious, time-consuming, and costly. Neural networks have emerged as a promising technique to automate brain lesion segmentation~\citep{ronneberger2015u,pereira2016brain}. However, training neural networks requires large amounts of well-annotated images to ensure good performance. The need for large training datasets limits the development of automatic brain lesion segmentation models since the scale of brain lesion datasets is often limited.

Data augmentation is a widely used technique to increase training dataset size and diversity in order to improve model performance. Conventional data augmentation strategies include random rotations, brightness adjustment, \etc Although these simple spatial and appearance transformations improve segmentation performance to some extent, they do not fundamentally increase the diversity of the dataset and provide a smaller boost in performance compared to acquiring new data. Recently, methods to perform data augmentation based on image fusion (such as Mixup~\citep{zhang2018mixup}, CutMix~\citep{yun2019cutmix}, \etc) have been developed to increase training dataset diversity. However, these approaches may shift the distribution of the original dataset~\citep{pinto2022using}, which is catastrophic for small datasets as the model learns non-representative features in the augmented images~\citep{zhang2018mixup}. Neural network approaches, such as Generative Adversarial Networks (GANs)~\citep{goodfellow2014generative} have been proposed to synthesize data for model training. However, the performance of GANs, similar to other neural networks, is constrained by the size of the training dataset, posing challenges in generating realistic images when only limited data is available. Regardless of the method used to create augmented samples, combining augmented and real samples does not guarantee the segmentation model will learn discriminative features to segment lesions on real samples as there is no supervision in the feature space. Therefore, we raise two questions: \textit{How can we generate realistic images that will not shift the original data distribution with limited training samples? How can we effectively use synthetic samples to train a segmentation model to perform well on real samples?}

We present a comprehensive framework to effectively augment brain imaging data with synthetic lesions to train a lesion segmentation model. Our framework has three stages: (1) train a lesion generator in a self-supervised manner, and sample feasible latent vectors from a constrained embedding space to create realistic paired lesion images and masks; (2) blend lesion images into existing brain images leveraging our novel image composition technique called Soft Poisson Blending (SPB); (3) train the segmentation model using a prototype consistency regularization term to align real and synthetic lesion features for better performance. The main contributions of our work are:
\begin{itemize}
\item We develop a data augmentation method to enhance the performance of neural networks trained to perform segmentation tasks. It comprises a two-stage adversarial autoencoder (AAE), consisting of shape and intensity generation networks, to generate on-the-fly new foreground regions when training. Distinct from other GANs, we designed a self-supervised approach where we simulate image-label pairs for training our AAE. This enables our AAE to learn a larger distribution of lesions and improve the quality of synthetic results.
\item We introduce Soft Poisson Blending (SPB), based on Poisson Image Editing~\citep{perez2023poisson}, to ensure realistic and smooth boundaries when inserting generated lesions into a brain image. SPB computes a refined guidance vector field that adjusts intensity values to make the synthetic lesion appear similar to the surrounding brain tissues.
\item We design a new loss term, prototype consistency regularization, to learn common features between synthetic and real training samples.
\end{itemize}

\begin{figure*}
\centering
\includegraphics[width=0.99\textwidth]{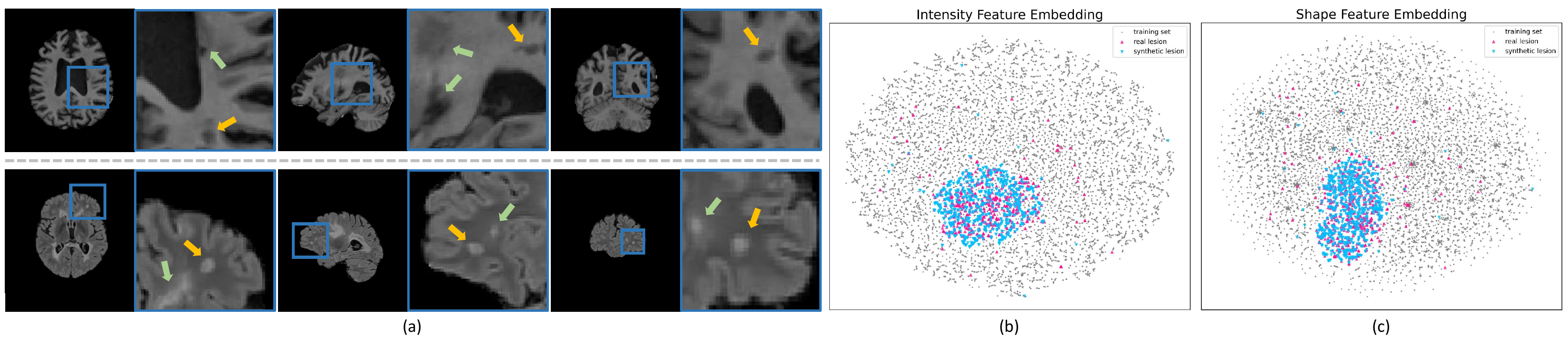}
\caption{(a) Real lesions (green arrow) and synthetic lesions (orange arrow) in images from the ATLAS v 2.0 (top) and MS Shift dataset (bottom) demonstrating the synthetic lesions have a similar appearance to real lesions. (b) t-SNE of the intensity embedding space for real lesions, synthetic lesions, and training samples. (c) t-SNE of the shape embedding space for real lesions, synthetic lesions, and training samples.}
\label{fig:gen_viz_with_tsne}
\end{figure*}

\section{Related Work}
\subsection{Data Augmentation for Data Scarcity}
To mitigate the problem of data scarcity when training neural networks, data augmentation techniques are used to increase the training dataset size thereby improving model performance. Conventional data augmentation techniques include shape transformations (flip, rotation, scaling, \etc) and appearance transformations (color jittering, brightness, and contrast adjustment, \etc)~\citep{krizhevsky2012imagenet,isensee2021nnu}. However, the diversity of the dataset achieved by such transformations is limited. Furthermore, the intrinsic characteristics of the training samples are not fundamentally changed, limiting improvements in model performance. The development of GANs enabled more advanced data augmentation where entirely new samples are created by the model. GANs can generate realistic samples for both natural~\citep{brock2018large} and medical~\citep{nie2017medical,schlegl2019f} images. GANs may also be designed to enable image-to-image translation, not generating images from noise, in order to create images of different modalities or characteristics. Nevertheless, GANs require large datasets when training the model to enable accurate image generation. Recently, image-manipulation-based data augmentation techniques~\citep{zhang2018mixup,yun2019cutmix,zhang2023carvemix} have been developed to increase training dataset size and variety by using a set of image manipulation rules to generate new samples. For instance, CutMix~\citep{yun2019cutmix} cuts and fuses patches from different images to create new training samples. Such techniques must be carefully designed and may shift the distribution of the training dataset especially when the original training dataset size is small~\citep{pinto2022using}. In this study, we combine the strengths of generative models and image-manipulation-based data augmentation techniques to synthesize realistic brain images with limited training samples.

\subsection{Poisson Blending in Deep Learning}
Poisson blending is an image processing technique to seamlessly integrate regions of a source image into a target image. In the context of data augmentation, this typically involves identifying the foreground of a source image and integrating this region into a target image. The blending process is guided by the gradient of the source image and the intensity of the target image~\citep{perez2023poisson}. Poisson blending generates more coherent images compared to simpler fusion methods such as Copy-Paste~\citep{ghiasi2021simple}. Poisson blending has been used for data augmentation for a variety of natural images~\citep{liu2021new}. In the context of medical imaging, \cite{tan2021detecting} developed a self-supervised learning strategy to detect abnormalities in chest X-ray images, and abnormalities were generated by fusing image patches into the target image using Poisson blending. \cite{wang2022anomaly} utilized Poisson blending to generate retinal images with lesions and different appearances than in the original training dataset to improve the performance of a lesion segmentation model. \cite{lee2023improving} introduced Poisson blending to the gastric disease classification task for better model generalization ability. However, all of these applications were applied to 2D images, and to our knowledge, Poisson blending has not yet been applied to 3D medical images.

\subsection{Prototype Learning}
Prototype networks~\citep{snell2017prototypical} were first presented for few-shot learning, where the network learns to symbolize each class using a prototype(\aka a representative vector in the embedding space). Each class prototype is typically obtained by computing the average feature of samples belonging to the class. This method was initially applied in image classification where distances between different class prototypes were maximized across training samples~\citep {snell2017prototypical}. Prototype learning was extended to image segmentation by computing cosine similarity between the class prototypes and individual pixel features. In this context for the test dataset pixels the class with which they have the highest similarity to its prototype. \citep{wang2019panet} designed a bidirectional prototype alignment mechanism for the few-shot image segmentation task. \citep{kuo2020featmatch} utilized class prototypes to augment training samples in the feature space. 
\citep{xu2022all} introduced a cyclic prototype consistency framework for semi-supervised medical image segmentation. In our work, we draw on the idea of prototype consistency to introduce a regularisation term during segmentation model training to align features between synthetic and real samples.

\section{Methodology}
\begin{figure*}[!htbp]
\centering
\includegraphics[width=0.95\textwidth]{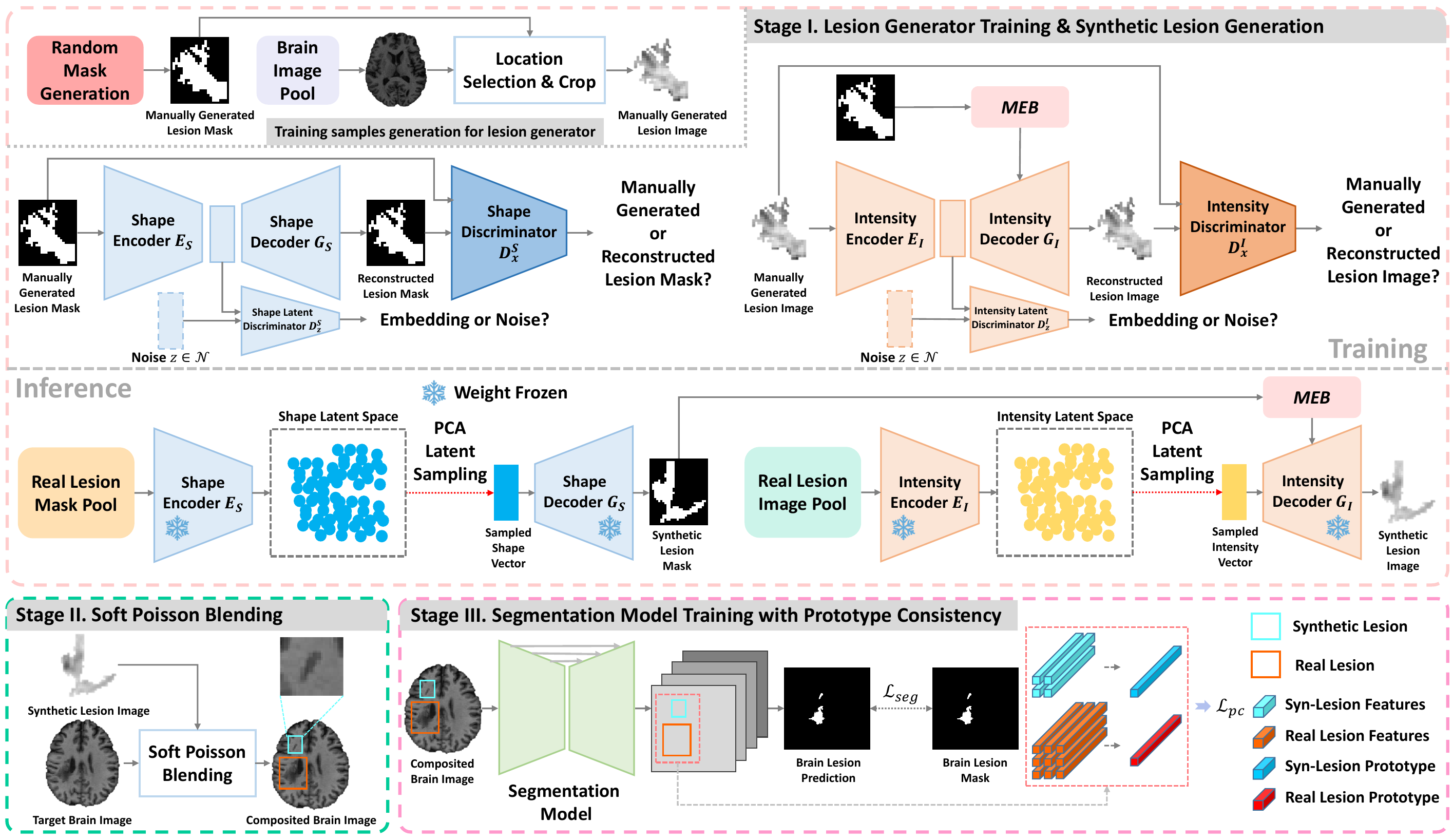}
\caption{Overview of our framework containing three stages. First, the lesion generator is trained via a self-supervised learning strategy and used to generate synthetic lesions based on constrained latent space sampling in Stage I. In Stage II, we seamlessly compose synthetic lesions into full brain images using the proposed Soft Poisson Blending (SPB) to increase the number of training samples. In Stage III, we train the downstream segmentation model with the prototype consistency regularization to align real and synthetic features.}
\label{fig:main_framework}
\end{figure*}

Fig.~\ref{fig:main_framework} illustrates our entire framework comprised of three stages: I. training the lesion generator, II. inserting synthetic lesions into brain images, and III. using the generated images for segmentation model training and incorporating prototype consistency regularization. Exemplar synthetic lesions and real lesions are shown in Fig.~\ref{fig:gen_viz_with_tsne} (a). The synthetic lesions have appearances similar to those of real lesions.

\subsection{Self-supervised AAE for Brain Lesion Synthesis}
Training a 3D generative model with a small dataset is challenging and achieving a good model performance is unlikely, therefore, we decompose brain lesion generation into two sub-tasks to reduce model complexity and use a self-supervised learning strategy during training. We designed two models: a shape adversarial auto-encoder (shape AAE) and an intensity adversarial auto-encoder (intensity AAE), to first create lesion masks and then perform texture synthesis to generate lesion images corresponding to the masks. Both shape and intensity AAEs are trained in a self-supervised manner. For lesion synthesis real lesion images are used to define the data distribution in the latent space and synthetic lesion images are created by sampling latent vectors from only this distribution before going through the decoder block of the trained AAEs.

\subsubsection{Shape and Intensity AAE Design}
\begin{figure}[!htbp]
\centering
\includegraphics[width=8.5cm]{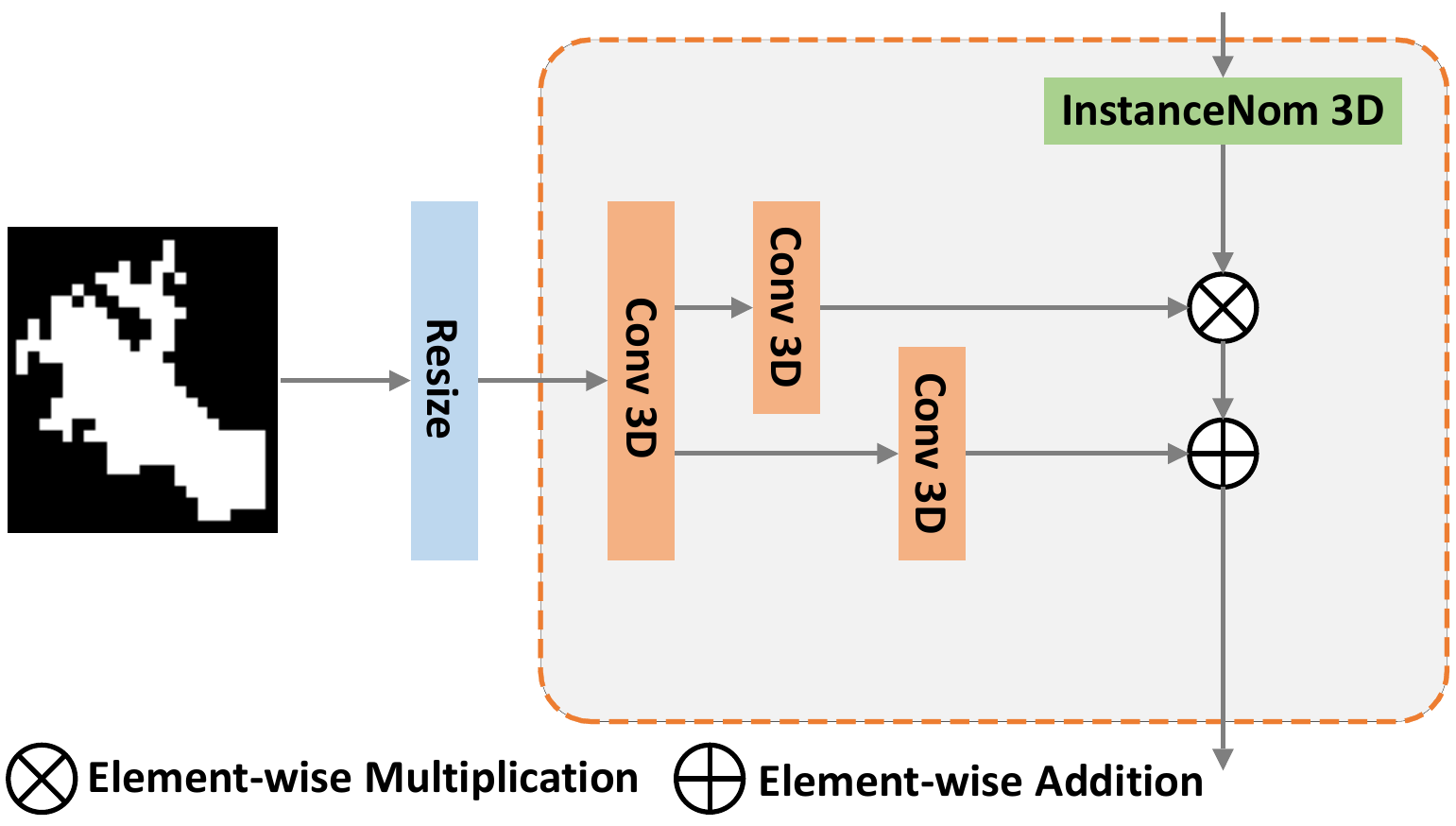}
\caption{The detailed structure of MEB block.}
\label{fig:meb}
\end{figure}

We follow the model architecture proposed in ~\citep{rombach2022high} to design the shape and intensity AAEs. As shown in Fig.~\ref{fig:main_framework}, each AAE contains an encoder $E$, a decoder $G$, an image discriminator $D_x$, and a latent discriminator $D_z$. Although the structures are similar, for the intensity AAE we introduce a mask embedding block (MEB)~\citep{huo2022brain} to provide shape guidance when generating the lesion intensity. The detailed structure of MEB is shown in Fig.~\ref{fig:meb}. It embeds the mask to the feature space to control the shape of synthetic lesions.

For training the AAE models we use a three-term loss: reconstruction loss $\mathcal{L}_{rec}$, latent adversarial loss $\mathcal{L}_{adv\_z}$, and image adversarial loss $\mathcal{L}_{adv\_x}$. $\mathcal{L}_{rec}$ computed as the mean absolute error (MAE) between the input image $\mathcal{I}$ and the reconstructed image $\mathcal{\hat{I}}$:
\begin{equation}
\mathcal{L}_{rec}=\|\mathcal{I} - \mathcal{\hat{I}}\|_{1}.
\end{equation}
$\mathcal{L}_{rec}$ guarantees $\mathcal{\hat{I}}$ and $\mathcal{I}$ look similar in general.

The latent adversarial loss $\mathcal{L}_{adv\_z}$ is formulated to ensure that the latent space of the lesions has a normal distribution:
\begin{equation}
\begin{aligned}
\mathcal{L}_{adv\_z}(D) &= \mathbb{E}[max(0,1+D_z(E(\mathcal{I})))] \\
&+ \mathbb{E}[max(0,1-D_z(\mathcal{N}(0,1)))],
\label{eq:loss_dz}
\end{aligned}
\end{equation}
\begin{equation}
\mathcal{L}_{adv\_z}(G) = -\mathbb{E}[D_z(E(\mathcal{I}))],
\label{eq:loss_gz}
\end{equation}
where $\mathcal{N}(0,1)$ is a normal distribution of mean $0$ and standard deviation $1$. Similarly, the image adversarial loss $\mathcal{L}_{adv\_x}$ is designed to ensure the reconstructed image is realistic in appearance. It is defined as:
\begin{equation}
\begin{aligned}
\mathcal{L}_{adv\_x}(D) &= \mathbb{E}[max(0,1+D_x(\mathcal{\hat{I}}) )] \\
&+ \mathbb{E}[max(0,1-D_x(\mathcal{I}))],
\label{eq:loss_dx}
\end{aligned}
\end{equation}
\begin{equation}
\mathcal{L}_{adv\_x}(G) = -\mathbb{E}[D_x(\mathcal{\hat{I}})].
\label{eq:loss_gx}
\end{equation}

\subsubsection{Training Set Generation}
\label{synthetic_lesion_generation_training_set_creation}
Unlike other generative models which train models with real image-mask pairs, we train our model in a self-supervised manner by generating lesion-mask pairs. As shown in Fig.~\ref{fig:gen_viz_with_tsne} (b) and (c), the latent distribution of real lesions (pink triangles) is a subset of the pre-generated lesions (grey dots), which indicates training on pre-generated lesion-mask pairs is sufficient to learn a representative feature embedding space for real lesion-mask pairs.

\begin{figure}[!htbp]
\centering
\includegraphics[width=8.5cm]{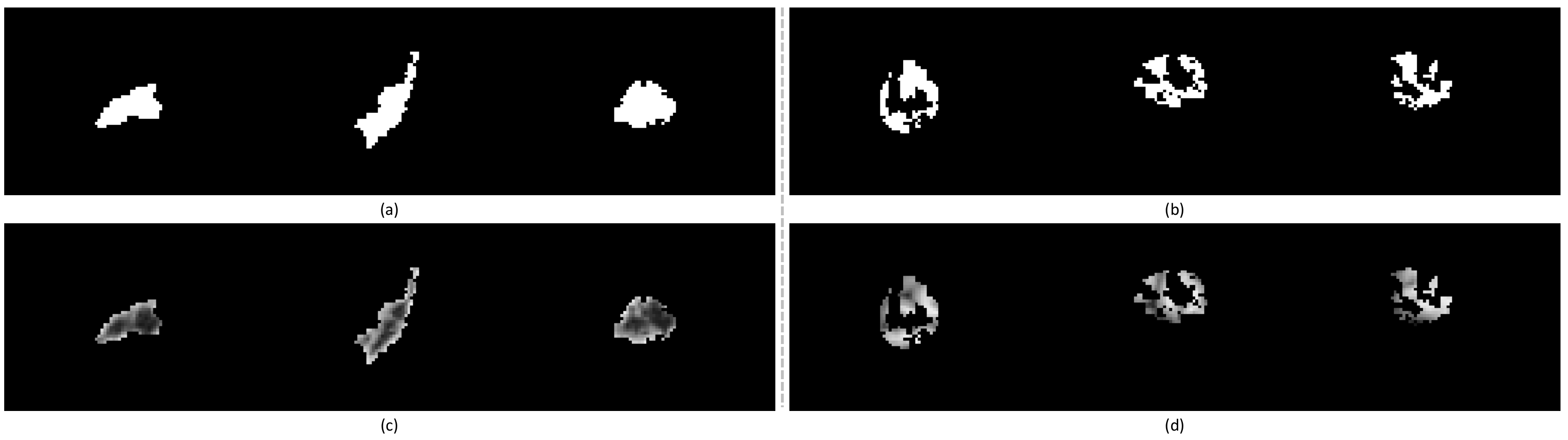}
\caption{ Lesion mask in three views for (a) real masks and (b) pred-generated masks for the shape AAE model training. Lesion images in three views for (c) real images and (d) pred-generated images for the intensity AAE model training.
}
\label{fig:lesion_and_mask_viz}
\end{figure}

The pre-generated lesion-mask pairs are created as follows. Inspired by~\citep{hu2023label}, we first generate $n | n \sim \mathcal{U}(1,5)$ ellipsoids with overlap to simulate a general lesion shape. The half-axis lengths of three directions follow the uniform distribution $\mathcal{U}(5,15)$. Elastic deformations controlled by $\sigma | \sigma \sim \mathcal{U}(3,6)$ are applied to the ellipsoids to make the shape more natural and irregular. Finally, we add Perlin noise~\citep{perlin1985image} to make the boundary more irregular. A comparison between real lesion masks and those generated by this approach is shown in Fig.~\ref{fig:lesion_and_mask_viz} (a) and (b). Note the complexity of the pre-generated masks exceeds that of real lesion masks, this enhances the AAE's ability to learn the reconstruction task.

To generate the lesion images we randomly select a location within the brain image from the training set and then extract the intensity values for voxels inside the pre-generated mask within that region. To increase variation in the training set, we apply the foreground intensity perturbation~\citep{huo2023arhnet} to randomly adjust the intensity values. Fig.~\ref{fig:lesion_and_mask_viz} (c) and (d) show real lesions and those generated by our approach. Note the styles of the images are similar, which ensures the pre-generated lesion-mask pairs are suitable for the model training.

\subsubsection{Constrained Sampling for New Lesion Synthesis}
Once AAE model training is complete, we freeze both shape AAE and intensity AAE model weights. As we see in Fig.~\ref{fig:gen_viz_with_tsne} (b) and (c), the latent space of real lesions is a subset of the latent space of the lesions generated for training the AAE models. Therefore, we use a constrained sampling strategy to synthesize lesions that are more similar to real lesions in the latent space. Specifically, we first obtain the latent embedding vectors for real lesion masks and lesion intensity images using the shape encoder $E_S$ and the intensity encoder $E_I$. Next, we apply Principal Component Analysis (PCA) to these vectors to obtain a dimensionality-reduced latent space representation. Note we apply PCA to the shape and intensity latent embedding vectors separately. We keep the top $K$ principal components for each space which cover $90\%$ of the latent embedding vector variance.

To create a new synthetic lesion we uniformly sample two vectors, one from the dimensionality-reduced shape latent space and one from the dimensionality-reduced intensity latent space. We then map these dimensionality-reduced latent vectors to the original latent embedding spaces. Finally, we use these two latent embedding vectors as the input to the shape decoder $G_S$ and intensity decoder $G_I$, respectively, to generate new synthetic lesion masks and images.

\subsection{Lesion Image Composition}
After we generate the synthetic lesion images, we have to fuse the lesion images with a brain image to generate training samples for the segmentation model training. We first select a plausible location in the brain for the generated lesion, then create a composite image using our modified Poisson image editing method called Soft Poisson Blending (SPB) to ensure the boundary between the image and lesion is seamless. We detail this approach below.

\subsubsection{Lesion Location Selection}
The brain has a regular anatomical structure, with distinct regions including the ventricle and brain stem. The location of brain lesions depends on the underlying pathology (\eg, stroke and multiple sclerosis). For the datasets in this work, we use FastFurfer~\citep{henschel2020fastsurfer} to segment the white matter area of the original patient's brain as a region proposal for lesion location selection. After obtaining the mask for white matter areas, we apply morphological binary erosion operation to shrink the masked region so that the synthetic lesion area will not exceed the mask boundary. We randomly (following a uniform distribution) select a voxel in the masked area as the center of the synthetic lesion. 

\subsubsection{Soft Poisson Blending}
We developed Soft Poisson Blending (SPB), which is a modified implementation of Poisson Image Editing~\citep{perez2023poisson}. The key idea of Poisson Blending is to use the Poisson partial differential equation under the Dirichlet boundary condition to specify the intensity value at the boundary area. We first adapt the conventional Poisson Blending approach to apply to 3D images. Second, we refine the guidance vector field, to ensure that the lesion foreground exhibits a natural internal appearance while having edge consistency with the background image.

\begin{figure}
\centering
\includegraphics[width=8.5cm]{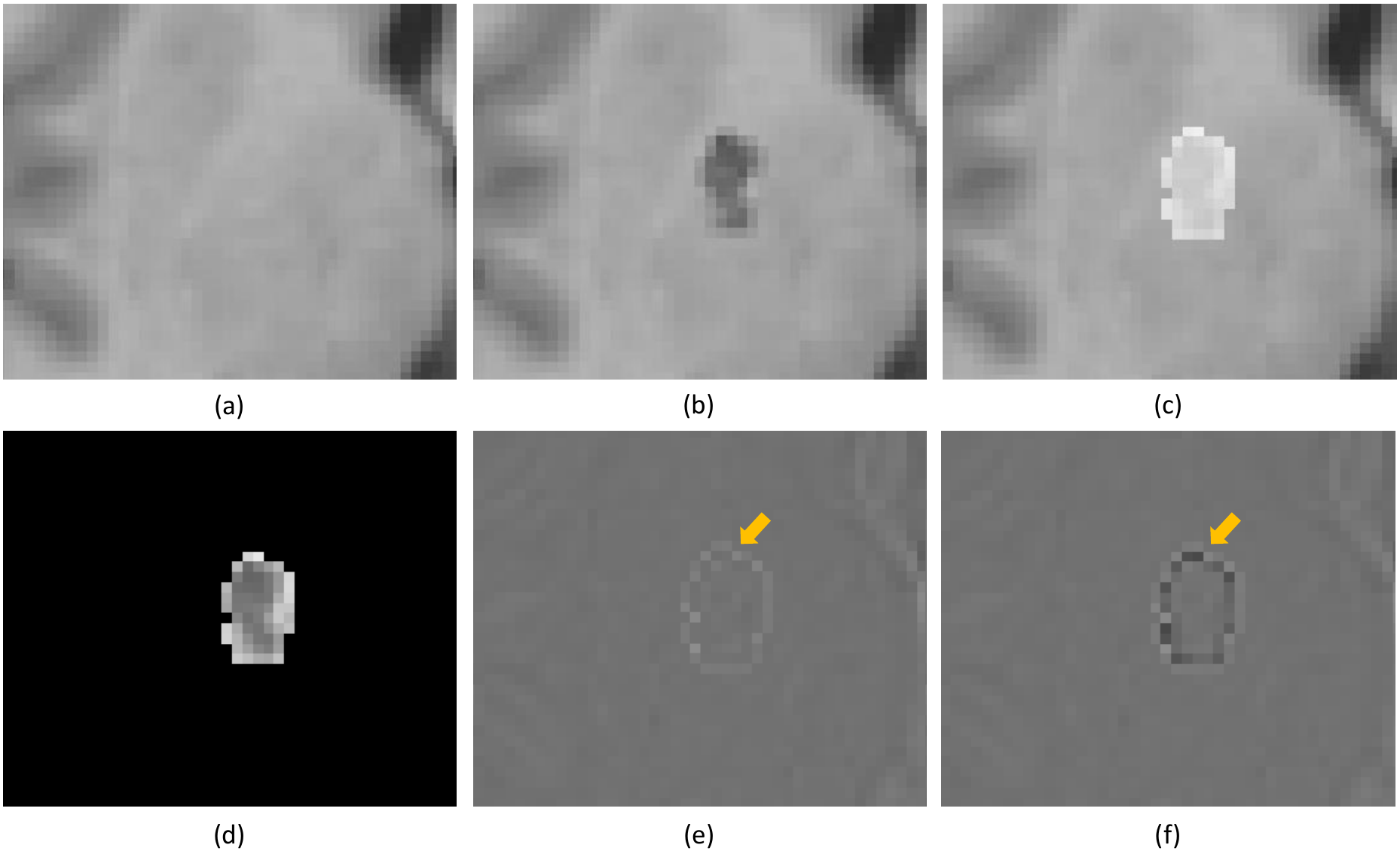}
\caption{(a) The target brain image used for the background image. (b) The composited image based on SPB. (c) The composited image based on the original Poisson Blending. (d) The synthetic brain lesion. (e) The guidance vector field used for SPB. (f) The guidance vector field used for the original Poisson Blending. The yellow arrow points to the gradient values on the region boundary.}
\label{fig:SPB_detail}
\end{figure}

For a brain image $s$, the target region that we would like to blend with the lesion image $g$, we define as $\Omega$. The boundary of $\Omega$ is defined as $\partial \Omega$. $f$, the value function defined on $\Omega$, has a value of $f^*$ at $\partial \Omega$. We solve the optimization problem defined as:
\begin{equation}
\min _f \iint_{\Omega}|\nabla f-V(x)|^2,\left.f\right|_{\partial \Omega}=\left.f^*\right|_{\partial \Omega},x\in\Omega
\label{eq:problem_formulation}
\end{equation}
where $V$ is the guidance vector field and $\nabla$ is the gradient operator. This equation guarantees that: i) The gradient of the foreground content is as close as possible to $V$. ii) The boundary pixel values of the foreground are consistent with the existing $s$, \ie a seamless transition. The solution under the Dirichlet boundary condition is the Poisson equation:
\begin{equation}
\Delta f=\operatorname{div} V(x),\left.f\right|_{\partial \Omega}=\left.f^*\right|_{\partial \Omega},x\in\Omega
\label{eq:poisson_equation}
\end{equation}
where $\Delta$ is the Laplacian operator, and $\operatorname{div}$ is the divergence operator. The guidance vector field $V$ (Fig.~\ref{fig:SPB_detail} (f)) is calculated as the mixed gradient of the brain image $s$ (Fig.~\ref{fig:SPB_detail} (a)), and the synthetic lesion image $g$ (Fig.~\ref{fig:SPB_detail} (d)) by selecting $(\nabla s,\nabla g)$. However, using this definition of $V$ to construct the blended image can lead to an unnatural appearance (Fig.~\ref{fig:SPB_detail} (c)). This unnatural appearance is caused by the absolute value of $\nabla g$ on $\partial \Omega$ being much higher than $\nabla s$ since for $g$ regions outside of the foreground are zero. Based on this observation, for Soft Poisson Blending, we modified the computation of $V$ as follows: 
\begin{equation}
\left.V(x)\right|_{x\in \Omega}= \begin{cases}\nabla s(x) & \left|\nabla s(x)\right|>|\nabla g(x)|\ \&\ x\in{\partial \Omega}, \\ \nabla g(x) & \text {otherwise.}\end{cases}
\label{eq:modified_guidance_vector_field}
\end{equation}
This results in the blended image becoming more realistic (Fig.~\ref{fig:SPB_detail} (e)) compared to the original Poisson Blending algorithm (Fig.~\ref{fig:SPB_detail} (f)).

\subsection{Prototype Consistency for Feature Alignment}
After synthetic lesions are blended into the brain images, the next step is training a segmentation model. Note that here the training dataset is a mixture of real and synthetic lesions which provides us with a unique opportunity to use this information to improve the lesion representations at the feature map level. We propose a consistency regularization, to prefer networks where feature maps for the two types of lesions (real and synthetic) are most similar. We hypothesize this will tend towards feature maps that are more general to the segmentation problem and less specific to features of the particular training dataset thereby increasing segmentation model robustness.

The feature map of real lesions is denoted as $F_{real}=F \cdot \mathds{1}\left[M=1\right]$, and the feature map of synthetic lesions is denoted $F_{syn}=F \cdot \mathds{1}\left[M=2\right]$. Here $F$ indicates the feature map of the composited image $\hat{I}$, and $\mathds{1}(\cdot)$ is an indicator function where the value is 1 if the condition is true, otherwise it is 0. Inspired by the prototypical network~\citep{snell2017prototypical}, we aim to force the segmentation model to learn similar feature distributions for $F_{real}$ and $F_{syn}$ via a class prototype. Specifically, we first obtain feature prototypes for both lesion types by averaging feature maps for the specific lesion type in the spatial dimension:
\begin{equation}
\mathcal{P}_{real}=\frac{\sum_{x,y,z} F^{(x,y,z)}_{real}}{\sum_{x,y,z} \mathds{1}\left[M^{(x,y,z)}=1\right]},
\end{equation}
\begin{equation}
\mathcal{P}_{syn}=\frac{\sum_{x,y,z} F^{(x,y,z)}_{syn}}{\sum_{x,y,z} \mathds{1}\left[M^{(x,y,z)}=2\right]},
\end{equation}
where $(x,y,z)$ is the spatial coordinate. The loss of the prototype differences can then be computed as:
\begin{equation}
\mathcal{L}_{pd}=\|\mathcal{P}_{real} - \mathcal{P}_{syn}\|_{1},
\end{equation}
where $\|\cdot\|_{1}$ is the L1-norm of a vector. 

Only optimizing $\mathcal{L}_{pd}$ neglects the intrinsic relation between class-specific features since it only minimizes the discrepancy between two class prototypes~\citep{asadi2023prototype}. To this end, we develop prototype relationship loss to maximize the consistency between relationship matrices constructed from the class prototype and class-specific features. Since the number of voxels in real and synthetic lesion areas can be different, we randomly sample $k$ feature vectors within each class to obtain $\mathcal{F}_{real} \in \mathbb{R}^{n\times c\times k}$ and $\mathcal{F}_{syn} \in \mathbb{R}^{n\times c\times k}$, where $c$ is the number of feature channels. To measure consistency we compute cosine similarity:
\begin{equation}
\cos(\mathcal{P},\mathcal{F})=\frac{\mathcal{P} \cdot \mathcal{F}}{\|\mathcal{P}\|_{2} \cdot \|\mathcal{F}\|_{2}},
\end{equation}
where $\|\cdot\|_{2}$ denotes the L2-norm. The prototype relationship loss is computed as:
\begin{equation}
\begin{aligned}
\mathcal{L}_{prd} &= \|\cos(\mathcal{P}^{i}_{real},\mathcal{F}^{i}_{real}) - \cos(\mathcal{P}^{i}_{real},\mathcal{F}^{i}_{syn})\|_{1} \\
&+ \|\cos(\mathcal{P}^{i}_{syn},\mathcal{F}^{i}_{real}) - \cos(\mathcal{P}^{i}_{syn},\mathcal{F}^{i}_{syn})\|_{1}.
\end{aligned}
\end{equation}

The loss function for training the segmentation model is:
\begin{equation}
\begin{matrix}
\mathcal{L}=\mathcal{L}_{seg} + \underbrace{\lambda_{1}\mathcal{L}_{pd} + \lambda_{2}\mathcal{L}_{prd}}_{\mathcal{L}_{pc}},
\end{matrix}
\end{equation}
where $\lambda_{1}$ and $\lambda_{2}$ are weight factors. $\mathcal{L}_{pc}$ is the prototype consistency, comprised of a difference and relationship loss term, and $\mathcal{L}_{seg}$ is the compound loss which comprises of the Dice and Cross-entropy loss functions.

\section{Experiments}
Our framework is evaluated on two public brain segmentation datasets: the ATLAS v2.0 dataset and the Shift MS segmentation dataset as described in Section~\ref{dataset_description}. We compare our approach to existing methods (Section~\ref{model_evaluation}) to show the superiority in both lesion synthesis and segmentation tasks. We further conduct ablation studies to validate the effectiveness of each component in our framework (Section~\ref{ablation_studies}).

\begin{table}[!htb]
\caption{Shift MS dataset details, including scanner location, type, magnetic field strength, and dataset split.} 
\centering
\vspace{2px}
\scalebox{0.78}{
\begin{tabular}{l|l|ll|cccc} 
\toprule
Dataset &Location &Scanner &Field &Trn &Dev$_{in}$ &Dev$_{out}$ &Evl$_{in}$ \\
\midrule
\multirow{4}*{MEESG-1} &Rennes              &S Verio   &3.0T &8  &2 &0 &5 \\  
                       &Bordeaux            &GE Disc   &3.0T &5  &1 &0 &2 \\  
                       &\multirow{2}*{Lyon} &S Aera    &1.5T &\multirow{2}*{10} &\multirow{2}*{2} &\multirow{2}*{0} &\multirow{2}*{17} \\
                       &                    &P Ingenia &3.0T &   &  &  &  \\
\midrule
ISBI                   &Best                &P Medical &3.0T &10 &2 &0 &9  \\
\midrule
PubMRI                 &Ljubljana           &S Mag     &3.0T &0  &0 &25 &0  \\
\bottomrule 
\end{tabular}}
\label{tab:dataset_msseg}
\end{table}

\subsection{Dataset and Preprocessing}
\label{dataset_description}
\subsubsection{ATLAS v2.0 Dataset}
The ATLAS v2.0 dataset~\citep{liew2022large} is a large stroke segmentation dataset that contains 655 T1-weighted brain images collected from 33 research cohorts. All images first have intensity standardization and are registered to the MNI-152 template (1mm$^{3}$ voxel spacing). A defacing step is applied to anonymize the scan. All lesion masks are annotated and then checked by two neurological experts. We randomly select 80\% of the dataset as the training set, and keep the remaining 20\% for model evaluation.

\subsubsection{Shift MS Dataset}
The Shift MS dataset~\citep{malinin2022shifts} is a multi-center white matter multiple sclerosis segmentation dataset comprised of \ie MSSEG-1~\citep{commowick2018objective}, ISBI\citep{carass2017longitudinal}, PubMRI~\citep{lesjak2018novel} and a private dataset collected from the University of Lausanne. We use the three public datasets for model training and evaluation since the private dataset is not publicly available. Detailed information on the dataset is shown in Table~\ref{tab:dataset_msseg}, the training set (Trn) is used for model training, the in-domain development set (Dev$_{in}$), the out-domain development set (Dev$_{out}$), and the in-domain evaluation set (Eva$_{in}$) are used for evaluating model segmentation performance only.

Each subject has two available modalities, FLAIR and T1 with contrast. In this work, we only use the FLAIR modality. All of the subjects have been preprocessed with image denoising, skull stripping, and bias field correction. We resample all images to 1mm$^3$ isotropic spacing. The ground-truth segmentation mask is determined by the consensus of annotations acquired from clinical experts.

\begin{table}[!htb]
\caption{Quantitative results of our method and other generative models for lesion synthesis. Here 'real' indicates models are trained using only real images, and 'synt' indicates models are trained with the self-supervised strategy.} 
\centering
\vspace{2px}
\scalebox{0.69}{
\begin{tabular}{l|c|c|cc|cc} 
\toprule
\multirow{2}*{Methods} &Training & \multirow{2}*{\#Param} &\multicolumn{2}{|c}{ATLAS v2.0} & \multicolumn{2}{|c}{Shifts MS} \\
&Data Type & &PSNR$\uparrow$ &MAE$\downarrow$ &PSNR$\uparrow$ &MAE$\downarrow$ \\
\midrule
AE                           &real &64.11M &31.72 &0.0014 &30.42 &0.0017 \\
\citep{rumelhart1986learning} &synt &64.11M &32.07 &0.0011 &31.56 &0.0014 \\
\midrule
AAE                            &real &64.41M &30.66 &0.0016 &31.33 &0.0014 \\
\citep{makhzani2015adversarial} &synt &64.41M &32.11 &0.0010 &32.84 &0.0009 \\
\midrule
f-AnoGAN            &real &78.15M &29.86 &0.0019 &28.35 &0.0023 \\
\citep{schlegl2019f} &synt &78.15M &30.22 &0.0017 &30.05 &0.0018 \\
\midrule
DDPM                    &real &74.23M &32.53 &0.0009 &32.38 &0.0009 \\
\citep{ho2020denoising} &synt &74.23M &34.48 &0.0007 &34.23 &0.0008 \\ 
\midrule
PNDM                    &real &74.23M &32.56 &0.0009 &32.32 &0.0009 \\
\citep{liu2022pseudo}   &synt &74.23M &34.87 &0.0007 &34.68 &0.0007 \\ 
\midrule
\multirow{2}*{Ours}     &real &67.08M &35.23 &0.0006 &34.38 &0.0007 \\
                        &synt &67.08M &\textbf{37.42} &\textbf{0.0004} &\textbf{36.57} &\textbf{0.0005} \\
\bottomrule 
\end{tabular}}
\label{tab:gen_exp}
\end{table}

\begin{table*}[!tbp]
\caption{Quantitative results of our method compared to other segmentation models for the Shift MS dataset. $*$ indicates that the p-value $<$ 0.05 compared to the second-best model performance computing with a paired Student's t-test.} 
\centering
\vspace{2px}
\scriptsize
\scalebox{0.95}{
\begin{tabular}{l|ccc|ccc|ccc} 
\toprule
\multirow{2}*{Methods} & \multicolumn{3}{|c}{Shifts MS Dev$_{in}$}  & \multicolumn{3}{|c}{Shifts MS Dev$_{out}$}  & \multicolumn{3}{|c}{Shifts MS Evl$_{in}$} \\
& DSC$\uparrow$ & ASD$\downarrow$ & HD95$\downarrow$ & DSC$\uparrow$ & ASD$\downarrow$ & HD95$\downarrow$ & DSC$\uparrow$ & ASD$\downarrow$ & HD95$\downarrow$ \\
\midrule
UNet~\citep{ronneberger2015u}             &62.73$\pm$21.04 &7.71$\pm$9.55 &20.40$\pm$18.91 &61.58$\pm$21.70 &4.34$\pm$8.57 &14.52$\pm$13.60 &51.74$\pm$20.29 &14.01$\pm$18.34 &29.02$\pm$23.83 \\ 
Attention UNet~\citep{oktay2018attention} &72.34$\pm$13.24 &2.46$\pm$2.77 &12.53$\pm$16.88 &64.80$\pm$22.98 &3.66$\pm$8.41 &12.83$\pm$15.65 &65.62$\pm$15.21 &3.34$\pm$5.07 &13.92$\pm$16.64 \\  
SwinUNETR~\citep{tang2022self}            &66.65$\pm$14.34 &3.54$\pm$4.51 &18.04$\pm$17.42 &61.39$\pm$22.29 &4.79$\pm$8.24 &15.27$\pm$14.60 &57.83$\pm$13.85 &5.38$\pm$8.24 &20.94$\pm$20.77 \\
MedNeXt~\citep{roy2023mednext}            &69.21$\pm$14.39 &10.93$\pm$12.96 &53.96$\pm$69.47 &64.99$\pm$22.99 &2.72$\pm$6.32 &14.97$\pm$17.89 &59.20$\pm$19.88 &21.54$\pm$37.39 &54.39$\pm$68.26 \\

\midrule
Ours &$\textbf{78.35$\pm$8.74}^*$ &\textbf{0.91$\pm$1.09} &$\textbf{8.15$\pm$12.06}^*$ &$\textbf{68.52$\pm$15.60}^*$ &$\textbf{1.73$\pm$3.24}^*$ &\textbf{12.26$\pm$19.37} &$\textbf{69.51$\pm$12.63}^*$ &$\textbf{1.60$\pm$2.17}^*$ &$\textbf{7.30$\pm$7.75}^*$ \\

\bottomrule 
\end{tabular}}
\label{tab:seg_exp_msseg}
\end{table*}

\subsection{Implementation Details}
The framework is implemented in PyTorch 1.13.1. All model training and validation were performed using an NVIDIA A100 40G GPU. For the lesion generator, the input mask and image size is $64\times 64\times 64$. We used the AdamW optimizer to train the shape and the intensity models, with the learning rate set to $1e-5$. The batch size was 4 and the total number of training epochs was 100. For the segmentation model, we use UNet~\citep{ronneberger2015u} as the backbone model. The input patch size is $128\times 128\times 128$ and the batch size is 2. We used the AdamW optimizer for the segmentation model training with the learning rate set to $1e-3$ and consecutively reduced with a cosine annealing strategy. A total of 500 epochs were set for the ATLAS v2.0 dataset and 1000 epochs for the Shift MS dataset. The loss function coefficients are set to $\lambda_1 = 1$, $\lambda_2 = 50$, for each dataset respectively, these values were chosen empirically.

\subsection{Model Evaluation}
\label{model_evaluation}
We evaluated our framework on (1) its ability to generate lesions, (2) the performance of the segmentation model trained with images generated using our framework compared to other models and (3) comparing our framework with other data augmentation techniques for training the segmentation model.

\subsubsection{Lesion Synthesis Performance}
To evaluate the effectiveness of our generative model and self-supervised training strategy, we compare the synthetic lesions generated by our framework to other generative models. We use the peak signal-to-noise ratio (PSNR) and mean absolute error (MAE) to evaluate synthetic results. Structural similarity (SSIM) is not suitable for this scenario because the large proportion of background dominates the small foreground area, leading to an inaccurate representation of the actual image quality. The measures are reported in Table~\ref{tab:gen_exp}. We compare our approach to both GAN and diffusion models. For each method, the model is trained either with only real lesions or only pre-generated image-mask pairs created using the self-supervised strategy (see Section~\ref{synthetic_lesion_generation_training_set_creation}). Across all models training with pre-generated data yields superior metrics compared to the real dataset, underscoring the ability of appropriate synthetic data to enhance model training. Additionally, our approach achieves the highest PSNR and lowest MAE, indicating its robustness and adaptability. Notably, the increased performance of our model is not attributable to the model's size, as evidenced by the comparison with f-AnoGAN which has the largest number of parameters but the lowest quantitative performance. The performance gains of our method are attributed to its innovative architecture and the utilization of the self-supervised training strategy.

\subsubsection{Downstream Segmentation Model Performance}
We compare the segmentation model performance using our framework to existing segmentation models. All models were trained from scratch. Note that our framework can use any backbone model to perform the segmentation task, here we consider UNet as the base segmentation model. Segmentation performance for all models was evaluated using the Dice similarity coefficient (DSC), average surface distance (ASD), and 95\% Hausdorff distance (HD95). Table~\ref{tab:seg_exp_atlas} shows the segmentation model performance for the ATLAS v2.0 dataset. Our approach consistently demonstrates improved performance for all metrics compared to all competing models, irrespective of the base model. This consistent overperformance is attributed to two pivotal enhancements: the use of an augmented brain lesion dataset, which closely mimics real-world appearance for model training, and the incorporation of the prototype consistency loss to align model features for real and synthetic lesions. These enhancements not only elevate the base model's ability to accurately segment images but also emphasize a crucial insight: the quality of the dataset plays a more critical role than the network architecture in achieving model generalization for segmentation tasks. Our approach, by leveraging high-fidelity synthetic data and strategic feature alignment, effectively unleashes the potential of classical segmentation models like UNet, establishing that dataset augmentation and targeted modifications to the training loss function can improve segmentation models' performances for brain stroke lesion segmentation.

\begin{table}[!htb]
\caption{Quantitative results of our method and other segmentation models for the ATLAS v2.0 dataset. $*$ indicates that the p-value $<$ 0.05 compared to the second-best model using a paired Student's t-test.} 
\centering
\vspace{2px}
\scriptsize
\scalebox{0.95}{
\begin{tabular}{l|ccc} 
\toprule
Methods & DSC$\uparrow$ & ASD$\downarrow$ & HD95$\downarrow$ \\
\midrule
UNet~\citep{ronneberger2015u}             &50.36$\pm$30.28 &20.25$\pm$25.46 &39.42$\pm$37.18 \\  
Attention UNet~\citep{oktay2018attention} &52.64$\pm$29.66 &19.75$\pm$23.70 &43.65$\pm$38.92 \\  
SwinUNETR~\citep{tang2022self}            &53.19$\pm$29.82 &21.47$\pm$51.44 &37.33$\pm$56.20 \\
MedNeXt~\citep{roy2023mednext}            &48.24$\pm$32.79 &21.67$\pm$52.65 &35.81$\pm$54.92 \\

\midrule
Ours &$\textbf{60.23$\pm$29.48}^*$ &$\textbf{6.32$\pm$13.68}^*$ &$\textbf{20.26$\pm$25.81}^*$ \\
\bottomrule 
\end{tabular}
}
\label{tab:seg_exp_atlas}
\end{table}

The quantitative metrics for model segmentation performance in the Shift MS Dataset are reported in Table~\ref{tab:seg_exp_msseg}. As with ATLAS v2.0, our framework has improved performance compared to the other models for the out-domain development set (Dev$_{out}$), which demonstrates our framework improves domain generalization ability.

\begin{figure*}[!htb]
\centering
\includegraphics[width=0.98\textwidth]{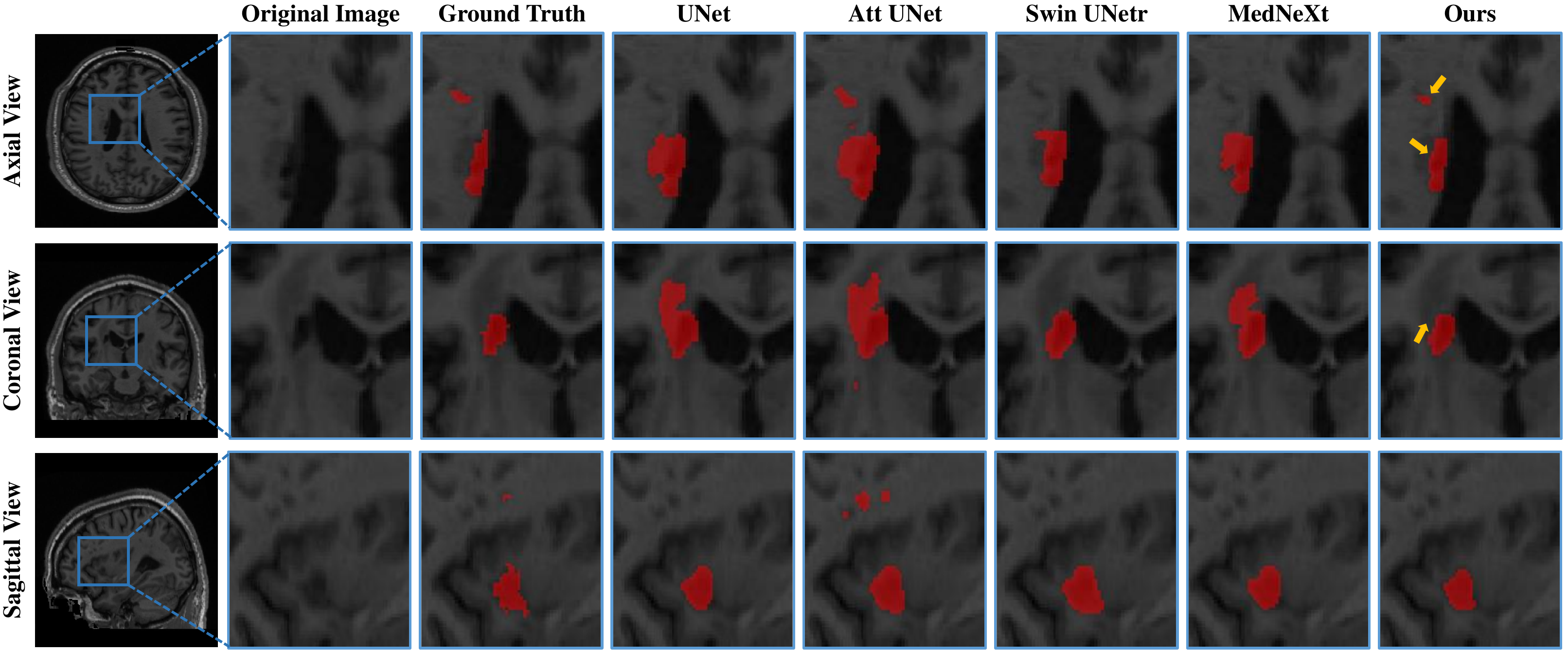}
\caption{Visualization of lesion segmentation in the ATLAS v2.0 dataset for different models. Here 'Att UNet' is short for the Attention UNet.}
\label{fig:atlas_seg_viz}
\end{figure*}

\begin{figure*}[!htb]
\centering
\includegraphics[width=0.98\textwidth]{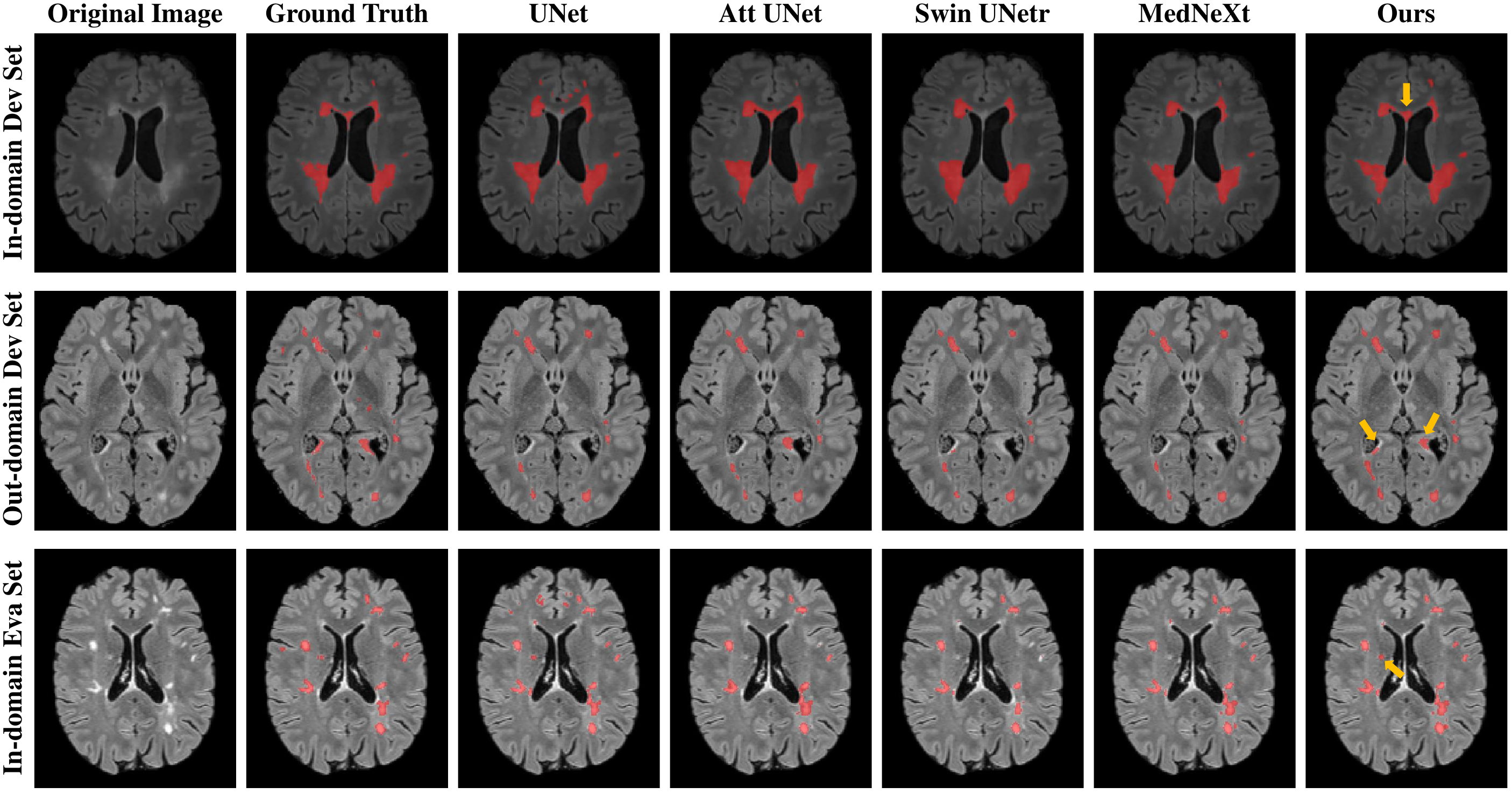}
\caption{Visualization of lesion segmentation on the Shift MS dataset for different models. Here 'Att UNet' is short for the Attention UNet.}
\label{fig:msseg_seg_viz}
\end{figure*}

Fig.~\ref{fig:atlas_seg_viz} and Fig.~\ref{fig:msseg_seg_viz} show qualitative segmentation results for the ATLAS v2.0 and Shift MS Dataset, respectively. These results demonstrate that our framework consistently provides more accurate predictions compared to other models. Notably, in regions indicated by the yellow arrows, other models have either false positives or fail to segment the foreground. The improvements in our framework are particularly evident in the segmentation of small lesions, which are often challenging for other segmentation models to recognize. 

\subsubsection{Comparisons with Different Data Augmentation Methods}
We compare the performance of our framework with other data augmentation methods, including voxel-based methods and GAN-based methods. Voxel-based data augmentation methods use a combination of two existing images to create a new synthetic image. Here we adopted six methods: \ie Mixup~\citep{zhang2018mixup}, CutMix~\citep{yun2019cutmix}, Copy-Paste~\citep{ghiasi2021simple}, TumorCP~\citep{yang2021tumorcp}, CarveMix~\citep{zhang2018mixup}, and SelfMix~\citep{zhu2022selfmix}. For the GAN-based methods~\citep{chaitanya2021semi}, a conditional GAN was adapted to generate either a deformation field (D), intensity field (I), or a combination of both (D+I) to change the structure and/or the appearance of images. Regardless of the data augmentation method, UNet was the segmentation model used. 

Quantitative segmentation model performance on the ATLAS v2.0 dataset and Shift MS dataset are shown in Table~\ref{tab:aug_exp_atlas} and Table~\ref{tab:aug_exp_msseg}, respectively. For the ATLAS v2.0 dataset, our framework achieves the best segmentation metric, improving DSC by 3.65, ASD by 5.99 mm, and HD95 by 3.27 mm compared to the second-best model. A similar trend is seen for the Shift MS dataset. One potential reason for the improvements seen in our framework is that the comparison voxel-based data augmentation methods create unrealistic foreground areas which may shift the decision boundary of the segmentation model and reduce their ability to generalize. For the GAN-based data augmentation methods, generated deformation and intensity fields do not change the intrinsic characteristics of the original images, which results in models that may overfit the training data.

\begin{table}[!htb]
\caption{UNet Segmentation model performance for our method and other data augmentation techniques on the ATLAS v2.0 dataset. For the GAN method, we consider deformation augmentation (D), intensity augmentation (I), and both (D+I). $*$ indicates that our framework significantly improves segmentation performance (p-value $<$ 0.05) compared to the second-best model using a paired Student's t-test.} 
\vspace{2px}
\scriptsize
\scalebox{0.85}{
\begin{tabular}{l|c|ccc} 
\toprule
Methods &Type &DSC$\uparrow$ &ASD$\downarrow$ &HD95$\downarrow$ \\
\midrule
Mixup~\citep{zhang2018mixup}        &Voxel &43.75$\pm$35.41 &23.45$\pm$20.48 &43.76$\pm$32.54 \\  
CutMix~\citep{yun2019cutmix}        &Voxel &47.32$\pm$32.46 &22.71$\pm$20.03 &40.35$\pm$31.56 \\  
Copy-Paste~\citep{ghiasi2021simple} &Voxel &55.24$\pm$31.42 &15.57$\pm$15.42 &25.78$\pm$27.44 \\
TumorCP~\citep{yang2021tumorcp}     &Voxel &55.64$\pm$32.59 &15.43$\pm$15.08 &24.86$\pm$26.94 \\
CarveMix~\citep{zhang2018mixup}     &Voxel &56.58$\pm$31.05 &12.31$\pm$16.54 &23.53$\pm$25.98 \\
SelfMix~\citep{zhu2022selfmix}      &Voxel &54.13$\pm$31.24 &16.78$\pm$17.55 &25.97$\pm$26.88 \\
\midrule

cGAN (D)~\citep{chaitanya2021semi}    &GAN &52.79$\pm$32.23 &20.43$\pm$22.48 &36.72$\pm$30.23\\
cGAN (I)~\citep{chaitanya2021semi}    &GAN &50.66$\pm$30.54 &21.14$\pm$23.84 &39.46$\pm$32.36 \\
cGAN (D+I)~\citep{chaitanya2021semi}  &GAN &51.48$\pm$31.26 &22.58$\pm$28.56 &37.22$\pm$31.28 \\
\midrule

Ours       &Mixed &$\textbf{60.23$\pm$29.48}^*$ &$\textbf{6.32$\pm$13.68}^*$ &\textbf{20.26$\pm$25.81} \\

\bottomrule 
\end{tabular}
}
\label{tab:aug_exp_atlas}
\end{table}

\begin{table*}[!htb]
\caption{UNet Segmentation model performance for our framework and other data augmentation techniques on the Shift MS dataset. For the GAN method, we consider deformation augmentation (D), intensity augmentation (I), and both (D+I). $*$ indicates that our framework significantly improves segmentation performance (p-value $<$ 0.05) compared to the second-best model using a paired Student's t-test.} 
\centering
\vspace{2px}
\scriptsize
\scalebox{0.90}{
\begin{tabular}{l|c|ccc|ccc|ccc} 
\toprule
\multirow{2}*{Methods} &\multirow{2}*{Type} & \multicolumn{3}{|c}{Shifts MS Dev$_{in}$}  & \multicolumn{3}{|c}{Shifts MS Dev$_{out}$}  & \multicolumn{3}{|c}{Shifts MS Evl$_{in}$} \\
& & DSC$\uparrow$ & ASD$\downarrow$ & HD95$\downarrow$ & DSC$\uparrow$ & ASD$\downarrow$ & HD95$\downarrow$ & DSC$\uparrow$ & ASD$\downarrow$ & HD95$\downarrow$ \\
\midrule
Mixup~\citep{zhang2018mixup}        &Voxel &52.64$\pm$32.68 &12.68$\pm$13.45 &26.75$\pm$28.46 &48.76$\pm$23.45 &13.54$\pm$14.23 &23.45$\pm$24.35 &38.45$\pm$21.56 &14.69$\pm$15.46 &38.58$\pm$39.43 \\  
CutMix~\citep{yun2019cutmix}        &Voxel &54.23$\pm$30.69 &11.54$\pm$12.53 &24.54$\pm$26.58 &50.43$\pm$22.59 &12.59$\pm$13.56 &21.76$\pm$22.69 &40.23$\pm$22.63 &13.42$\pm$12.53 &36.44$\pm$37.22 \\  
Copy-Paste~\citep{ghiasi2021simple} &Voxel &60.36$\pm$15.48 &8.42$\pm$9.64 &22.43$\pm$25.46 &58.46$\pm$19.53 &11.23$\pm$10.77 &20.69$\pm$21.34 &43.56$\pm$23.28 &10.85$\pm$11.49 &25.12$\pm$26.41 \\  
TumorCP~\citep{yang2021tumorcp}     &Voxel &61.55$\pm$13.55 &6.46$\pm$7.69 &20.15$\pm$27.33 &62.44$\pm$18.46 &9.53$\pm$6.75 &18.43$\pm$20.67 &51.12$\pm$20.51 &9.34$\pm$8.32 &18.32$\pm$15.46 \\  
CarveMix~\citep{zhang2018mixup}     &Voxel &65.28$\pm$12.47 &5.24$\pm$5.53 &15.63$\pm$18.42 &63.86$\pm$17.63 &8.75$\pm$7.53 &15.23$\pm$20.49 &54.53$\pm$18.24 &6.43$\pm$7.22 &15.44$\pm$12.75 \\  
SelfMix~\citep{zhu2022selfmix}      &Voxel &62.45$\pm$14.62 &7.53$\pm$6.45 &19.44$\pm$20.47 &60.54$\pm$18.45 &10.84$\pm$7.53 &16.29$\pm$21.54 &52.36$\pm$19.42 &7.52$\pm$6.31 &16.32$\pm$11.42 \\  
\midrule

cGAN (D)~\citep{chaitanya2021semi}   &GAN &62.87$\pm$22.54 &9.53$\pm$7.65 &19.78$\pm$22.23 &62.12$\pm$20.68 &6.52$\pm$7.98 &17.32$\pm$21.34 &52.65$\pm$18.23 &12.89$\pm$10.25 &28.36$\pm$24.35 \\
cGAN (I)~\citep{chaitanya2021semi}   &GAN &60.23$\pm$25.65 &10.54$\pm$8.33 &23.45$\pm$24.25 &60.98$\pm$22.45 &8.51$\pm$8.21 &18.55$\pm$20.35 &51.25$\pm$18.78 &13.45$\pm$11.54 &30.48$\pm$26.35 \\
cGAN (D+I)~\citep{chaitanya2021semi} &GAN &61.27$\pm$24.47 &9.23$\pm$8.11 &22.36$\pm$21.69 &61.73$\pm$21.59 &6.78$\pm$8.07 &17.21$\pm$19.24 &52.73$\pm$17.63 &12.45$\pm$10.48 &26.45$\pm$24.51\\
\midrule

Ours       &Mixed &$\textbf{78.35$\pm$8.74}^*$ &$\textbf{0.97$\pm$1.09}^*$ &$\textbf{8.15$\pm$12.06}^*$ &$\textbf{68.52$\pm$15.60}^*$ &$\textbf{1.73$\pm$3.24}^*$ &\textbf{12.26$\pm$19.37} &$\textbf{69.51$\pm$12.63}^*$ &$\textbf{1.60$\pm$2.17}^*$ &$\textbf{7.30$\pm$7.75}^*$ \\  

\bottomrule 
\end{tabular}}
\label{tab:aug_exp_msseg}
\end{table*}

\subsection{Ablation Studies}
\label{ablation_studies}
We validate the effectiveness of the three modules in our framework \ie Lesion synthesis (SL), Soft Poisson Blending (SPB), and Prototype Consistency (PC). All ablation studies were conducted on the ATLAS v2.0 dataset and Shift MS dataset.

\subsubsection{Synthetic Lesion} We validated the performance of the segmentation model when using only synthetic lesions for training. The results are shown in the first row of Table~\ref{tab:ablation_atlas} and Table~\ref{tab:ablation_msseg} for the ATLAS v2.0 and Shift MS datasets, respectively. For the ATLAS v2.0 dataset, training with only synthetic lesions improves segmentation performance compared to using only the real data for model training. In contrast, segmentation performance on the Shift MS dataset is worse. The reason for this discrepancy we believe is dataset size, the ATLAS v2.0 dataset contains 655 images but the Shift MS dataset only has 33 images. While foreground mismatch and boundary artifacts caused by directly inserting the synthetic lesions can increase the model's generalization, they are catastrophic for a small dataset like the MS Shift dataset since the synthetic lesions with boundary artifacts will shift the segmentation model feature distribution.

\subsubsection{Soft Poisson Blending} Applying SPB to achieve a consistent appearance with real lesions and a seamless boundary improves the segmentation model performance for both datasets (the second row of Table~\ref{tab:ablation_atlas} and Table~\ref{tab:ablation_msseg}). Our results demonstrate that resolving the inconsistent appearance of synthetic lesions improves the model performance. 

\subsubsection{Prototype Consistency}
To address the potential feature gap caused by synthetic and real images, we introduced prototype consistency regularization. This penalty, applied to both real and synthetic lesions, ensures the segmentation model learns similar features for lesions regardless of origin. Results shown in the third row of  Tables~\ref{tab:ablation_atlas} and~\ref{tab:ablation_msseg} demonstrate applying prototype consistency regularization solely to synthetic lesions yields improved segmentation model performance. Moreover, integrating this regularization with a consistent lesion appearance further enhances segmentation performance, as evidenced in the fourth row of Table~\ref{tab:ablation_atlas}. The Shift MS dataset (Table~\ref{tab:ablation_msseg}) demonstrates a substantial improvement in segmentation performance compared to models where the consistency penalty was not employed highlighting that feature alignment is most important for small datasets where even a small shift in the synthetic lesion distribution can affect segmentation performance.

\begin{table}[!htb]
\caption{Ablation study on the components of our framework: synthetic lesions (SL), Soft Poisson Blending (SPB), and prototype consistency loss (PC) for the ATLAS v2.0 dataset. UNet is the segmentation model.} 
\vspace{2px}
\scalebox{0.92}{
\begin{tabular}{ccc|ccc} 
\toprule
SL &SPB &PC & DSC$\uparrow$ & ASD$\downarrow$ & HD95$\downarrow$ \\
\midrule
$\checkmark$ & & &54.86$\pm$29.84 &16.96$\pm$24.59 &34.98$\pm$36.01 \\  
$\checkmark$ &$\checkmark$ & &58.71$\pm$27.20 &11.30$\pm$17.57 &27.17$\pm$30.28 \\
$\checkmark$ & &$\checkmark$ &59.38$\pm$30.80 &18.93$\pm$72.62 &30.00$\pm$73.03 \\  
$\checkmark$ &$\checkmark$ &$\checkmark$ &\textbf{60.23$\pm$29.48} &\textbf{6.32$\pm$13.68} &\textbf{20.26$\pm$25.81} \\

\bottomrule 
\end{tabular}}
\label{tab:ablation_atlas}
\end{table}

\begin{table}[!htb]
\caption{Ablation study on the components of our framework: synthetic lesions (SL), Soft Poisson Blending (SPB), and prototype consistency loss (PC) for the Shift MS dataset. UNet is the segmentation model.} 
\vspace{2px}
\scalebox{0.92}{
\begin{tabular}{ccc|ccc} 
\toprule
\multirow{2}*{SL} &\multirow{2}*{SPB} & \multirow{2}*{PC} & \multicolumn{3}{|c}{Shifts MS Dev$_{in}$} \\
& & & DSC$\uparrow$ & ASD$\downarrow$ & HD95$\downarrow$ \\
\midrule
$\checkmark$ & & &59.54$\pm$24.61 &9.68$\pm$12.64 &22.36$\pm$17.31 \\  
$\checkmark$ &$\checkmark$ & &67.16$\pm$14.26 &5.98$\pm$6.74 &20.06$\pm$16.99 \\
$\checkmark$ & &$\checkmark$ &70.05$\pm$20.63 &3.11$\pm$4.39 &13.49$\pm$17.06 \\  
$\checkmark$ &$\checkmark$ &$\checkmark$ &\textbf{78.35$\pm$8.74} &\textbf{0.97$\pm$1.09} &\textbf{8.15$\pm$12.06} \\
\hline
\hline
\multirow{2}*{SL} &\multirow{2}*{SPB} & \multirow{2}*{PC} & \multicolumn{3}{|c}{Shifts MS Dev$_{out}$} \\
& & & DSC$\uparrow$ & ASD$\downarrow$ & HD95$\downarrow$ \\
\midrule
$\checkmark$ & & &58.48$\pm$25.06 &6.89$\pm$9.77 &17.75$\pm$15.71 \\  
$\checkmark$ &$\checkmark$ & &59.23$\pm$26.13 &7.11$\pm$10.79 &16.36$\pm$16.76 \\
$\checkmark$ & &$\checkmark$ &59.99$\pm$24.63 &3.30$\pm$7.16 &15.88$\pm$17.47 \\  
$\checkmark$ &$\checkmark$ &$\checkmark$ &\textbf{68.52$\pm$15.60} &\textbf{1.73$\pm$3.24} &\textbf{12.26$\pm$19.37} \\
\hline
\hline
\multirow{2}*{SL} &\multirow{2}*{SPB} & \multirow{2}*{PC} & \multicolumn{3}{|c}{Shifts MS Evl$_{in}$} \\
& & & DSC$\uparrow$ & ASD$\downarrow$ & HD95$\downarrow$ \\
\midrule
$\checkmark$ & & &42.48$\pm$23.76 &18.42$\pm$19.71 &33.52$\pm$23.88 \\  
$\checkmark$ &$\checkmark$ & &53.89$\pm$22.22 &18.20$\pm$28.73 &41.92$\pm$54.82 \\
$\checkmark$ & &$\checkmark$ &54.55$\pm$25.35 &6.05$\pm$7.33 &17.30$\pm$15.15 \\  
$\checkmark$ &$\checkmark$ &$\checkmark$ &\textbf{69.51$\pm$12.63} &\textbf{1.60$\pm$2.17} &\textbf{7.30$\pm$7.75} \\
\bottomrule 
\end{tabular}}
\label{tab:ablation_msseg}
\end{table}

\section{Conclusions \& Discussions}
We presented a comprehensive framework to augment existing training samples for brain lesion segmentation via a two-stage adversarial autoencoder (AAE) to generate new lesion images.  The AAE is trained in a self-supervised manner, but generates synthetic lesions with the same latent space distribution as real lesions. We then augment the synthetic images by using Soft Poisson Blending (SPB) to create a seamless boundary between foreground and background, eliminating boundary artifacts. Finally we introduce a prototype consistency regularisation term during segmentation model training to ensure similar features across synthetic and real lesions are learnt. The synthetic lesion samples boost segmentation model performance under the supervision of the prototype consistency penalty. Experiments on two public datasets demonstrate that our framework outperforms other data augmentation approaches and methods that only adapt augmented samples for model training. We do not compare our approach to models based on pre-trained datasets such as SAM~\citep{kirillov2023segment} because they are pre-trained on a large-scale dataset, making direct comparisons unfair. Besides, SAM-based methods largely depend on accurate user prompts to achieve good segmentation results, which differs from our fully automatic setting requiring no prompt. Currently, our framework is validated on brain lesion MRI datasets. Extending our framework to other image modalities and other organs will be future work. Additionally, we will explore adding conditions to further control the process of lesion image synthesis for controllable data augmentation in the future.

\section*{Acknowledgments}
This work was supported by Centre for Doctoral Training in Surgical and Interventional Engineering at King's College London; the funding from the Wellcome Trust Award (218380/Z/19/Z) and the Wellcome/EPSRC Centre for Medical Engineering (WT203148/Z/16/Z).

\bibliographystyle{model2-names.bst}\biboptions{authoryear}
\bibliography{refs}

\end{document}